\documentclass[reprint, amsmath, amssymb, aps, pra]{revtex4-2}
\usepackage{graphicx}% Include figure files
\usepackage{bm}% bold math
\usepackage{hyperref}% add hypertext capabilities
\usepackage{tabularx}

\begin{document}

\title{Quench-induced spontaneous currents in rings of ultracold fermionic atoms}

\author{Daniel G. Allman}
\author{Parth Sabharwal}
\author{Kevin C. Wright}
\affiliation{Department of Physics and Astronomy, Dartmouth College, 6127 Wilder Laboratory, Hanover NH 03766, USA}

\begin{abstract}
We have observed the spontaneous appearance of currents in a ring of ultracold fermionic atoms ($^6$Li) with attractive interactions, following a quench to a BCS-like pair superfluid. We have measured the winding number probability distribution for a range of quench rates, with a quench protocol using simultaneous forced evaporation and interaction ramps to achieve faster effective quench rates with less atom loss than a purely evaporative quench. We find that for the fastest quenches the mean square winding number of the current follows a scaling law in the quench rate with exponent $\sigma=0.24(2)$, which is somewhat lower than that predicted by the Kibble-Zurek mechanism (KZM) for the three-dimensional XY model (1/3), and unexpectedly closer to the value obtained from mean-field theory (1/4). For slower quenches non-universal effects become significant, and we observe a lower rate of spontaneous current formation that does not follow a simple scaling law.
\end{abstract}

\maketitle

\section{Introduction}
When a system is quenched through a second order phase transition, the finite speed at which information propagates from one region to another prevents the order parameter from taking on a uniform global value. Local fluctuations in the disordered phase can cause the order parameter to take on independent values in different regions of space as it grows after the system crosses the phase transition~\cite{Kibble1976}. As these domains grow and merge, the variations between them can lead to the formation of topological features (e.g. vortex lines). The formation and subsequent evolution of these topological features over short timescales can be quite complex, but over longer timescales remarkable macroscopic effects can emerge, such as the appearance of a stable quantized current in a multiply-connected system that was non-circulating before the quench.~\cite{ZurekCosmologicalN85}. 

The probability for a spontaneous current to appear in the ordered phase after a quench is related to the average number of independent domains expected to form around a closed path. The core assertion of the Kibble-Zurek mechanism (KZM) is that the initial density of these domains should scale with quench rate, with an exponent that is determined by the universal static properties of the phase transition. There have been numerous experimental observations of spontaneous defect formation in different condensed matter systems following a quench, including liquid crystals~\cite{Chuang1991, Bowick1994} superfluid $^4$He~\cite{Hendry1993, Hendry1994, Dodd1998, Dodd1999} and $^3$He~\cite{Bauerle1996, Ruutu1996}, superconductors~\cite{MonacoPRB2009}, ion crystals~\cite{Ulm2013, Pyka2013}, and ultracold quantum gases of bosonic~\cite{Sadler2006,Saito2007, SchererVortexPRL07, WeilerSpontaneousN08, Navon2010, Lamporesi2013, Corman2014, NavonScience2015, DonadelloPRA2016, BeugnonIOP2017, GooPhysRevLett2022, RabgaPRA2023}, and fermionic~\cite{Ko2019, DykePRL2021, Liu2021, LeeArXiv2023} atoms. There is consistently strong qualitative support for the predictions of the KZM in these experiments, but establishing clear quantitative agreement between experimental observations and the ideal version of the theory has often been difficult.

There are many reasons for these experimental challenges, from defect dynamics to the implementation of the quench, and most of them involve physics that is interesting in its own right. We will focus here on issues most relevant to experiments like our own; a broader review of these topics can be found elsewhere~\cite{delCampoJMPA2014}. One major practical consideration is that the defects take time to form, continue to evolve, and the defect density generally decreases with time due to annihilation and decay. This fact is especially important in liquid helium, where the most carefully performed experiments to date measured vortex densities two orders of magnitude lower than expected, with the discrepancy attributed to rapid decay.~\cite{Dodd1998, Hendry2000}. Efforts to measure scaling with quench rate are also complicated by the fact that annihilation rates increase with defect density, leading to saturation of defect density for fast quenches in some experiments~\cite{Ko2019,GooPRL2021,RabgaPRA2023}.

Zurek anticipated these difficulties in his original proposal and suggested that measuring the current around an annulus long after a quench could provide a more stable measure of the initial defect density. He also predicted a linear relationship between the defect density and the variance of the winding number in a sufficiently large ring when the defects are spatially well-defined (See also~\cite{XiaPRD2020}). Other challenges have prevented a successful quenched-ring experiment with liquid helium, but experiments with multiply-connected superconductors~\cite{MoncacoPRL2006,AaroeIEEE2007,MonacoPRB2009,WeirIOP2013} and ultracold Bose gases~\cite{Corman2014} have established the feasibility of this approach. 

This paper is organized as follows: In Sec.~\ref{sec:ideal_v_real}, we discuss the application of the KZM to our experimental conditions and explain challenges in making quantitative comparisons with predictions of the KZM in various settings. In Sec.~\ref{sec:exp_methods} we describe our experimental procedures for preparing, quenching, and measuring the current, and provide information about the state of the system during the quench. In Sec.~\ref{Results and Discussion}, we present the results of our experiments, discuss a few key observations pertaining to these results, then give a brief overview of potential extensions to this work. Section~\ref{PKZM} offers a simple theoretical description of the quench dynamics and statistics from a mean-field stochastic Landau-Ginzburg model to provide further context to the discussion, before we summarize our findings in Sec.~\ref{conclusion}.

\section{Comparing Ideal and Real Systems}\label{sec:ideal_v_real}

When the reduced temperature $\epsilon=1-T/T_c$ of a homogeneous system is linearly swept across a second order transition [$\epsilon(t)=t/t_q$ with $1/\dot{\epsilon}=t_q$ the quench time], the KZM predicts the typical size of the uncorrelated domains that form, $\hat{\xi}_{KZ}\sim t_q^{\nu/(1+\nu z)}$. Within the so-called adiabatic-impulse picture, this domain size is obtained from the equilibrium correlation length $\xi\sim|\epsilon|^{-\nu}$ at the moment $t= -\hat{t}_{KZ}\sim -t_q^{\nu z/(1+\nu z)}$ the order parameter ceases to follow the quench adiabatically due to the diverging relaxation time $\tau\sim\xi^z$. These independent phase domains are ``frozen in" until $t=+\hat{t}_{KZ}$ after the transition when the order parameter can again follow the quench adiabatically. Around this time the rapidly growing condensate permits topological order to form from the merging independent phase domains.

For a thin one-dimensional ring of circumference $C$ with domain size $\hat{\xi}_{KZ}\ll C$, the number of uncorrelated domains formed at the freeze-out point is $N_d\sim C/\hat{\xi}_{KZ}$. This is the scenario originally envisaged by Zurek~\cite{ZurekCosmologicalN85}, who further predicted random-walk scaling of the mean absolute winding number $\langle|w|\rangle\sim N_d^{1/2}$ in accordance with a true Gaussian distribution of the winding numbers, thus connecting the measurable $\langle|w|\rangle$ with the exponents $\nu$ and $z$. For purposes of comparison with theory, it is important to note that the scaling exponent of the variance of the winding number is the same as the scaling exponent for the defect density. While a fairly robust power law scaling of $\langle w^2\rangle$ with $N_d$ is expected and has indeed been observed for even the modest defect densities achieved in prior annular Bose gas experiments~\cite{Corman2014, Aidelsburger2017}, the Zurek or random walk regime of Gaussian probabilities is only accessed when $\langle w^2\rangle\sim\langle|w|\rangle^2$, which has not yet been realized in this setting.

While the KZM elegantly predicts scaling laws for the defect density for an idealized set of conditions, it overlooks several important practical considerations. The first is that controlling the parameters that drive the phase transition without introducing unwanted effects on the system is an experimentally non-trivial task. Secondly, the dynamics that lead to topological ordering from the merging domains involve complicated stochastic processes, even if the initial domains are well-conditioned~\cite{Aidelsburger2017, CarmiPRL2000}. As noted above, the defects also continue to evolve and are subject to decay and annihilation processes even when the order parameter is well-established. It is then a natural question to ask when the right time to measure the defect density is, and this has been explicitly addressed theoretically~\cite{CheslerPRX2015} and in several experiments~\cite{NavonScience2015, BeugnonIOP2017}.

In a non-uniform system such as a quantum gas in a harmonic trap, the KZ scaling exponents are modified, as causality restricts the trap region in which the KZM can proceed as usual. There has been some notable success adapting the basic KZM framework to describe inhomogeneous conditions by including finite-size corrections to correlation functions and accounting for modified causality conditions in non-uniform systems~\cite{MonacoPRB2009, WeirIOP2013, del_CampoPRA2011}. In harmonic and uniform box traps, post-quench dynamics of the spontaneously nucleated vortices lead to saturation of the defect density for fast quenches. These observations complicate interpretations of the KZ scaling laws in those settings. KZM studies in ring-shaped traps have some advantages for mitigating these effects: The ring is azimuthally homogeneous, and the spontaneous excitations (persistent currents) are much longer lived than vortices in a simply-connected trap.

Because the expected current scaling exponents are usually sub-unity, experimental efforts to accurately measure them are hampered by practical limits on both system size and maximum achievable quench rates. The latter, for an ultracold atomic gas, are typically set by the timescale for thermalization, of order $h/\mu$ with $h$ Planck's constant and $\mu$ the chemical potential \cite{Corman2014}. Thus, fast quenches with a large dynamical range are desirable for precise measurements of the scaling exponent in this regime. In our experiments, the use of $^6$Li allows for rapid quenches (in comparison to those performed using weakly interacting bosons) to be enacted by leveraging both large local thermalization rates and fast ramps of the interaction strength by varying the magnetic field around a Feshbach resonance. In particular, the typical elastic collision rates in the regions of highest density in our trap are several tens of kHz, substantially greater than the several hundred Hz collision rates typical of weakly interacting boson experiments. Furthermore, the timescale with which the scattering length can adjust to changes in the external Feshbach field is limited only by the extremely rapid Landau-Zener dynamics of scattering pairs~\cite{Ketterle2008Springer} or excitation of the Higgs mode~\cite{Behrle2018,Barresi2023}, with the more practical limitation being technical challenges to enacting rapid magnetic field changes. 

Another important consideration in conducting KZM experiments with quantum gases is that system sizes are comparatively limited, and finite-size effects are not negligible. In the case of slower quenches performed in a ring, $\hat{\xi}_{KZ}\rightarrow C$, and the concept of a  correlation length becomes ill-defined as correlations begin to extend around the entire ring. Theoretical analyses in this limit predict a doubling of the scaling exponent, an exponential damping of the winding number variance, and winding number statistics that become dependent on the exact phase profile within each domain \cite{MonacoPRB2009,ZurekIOP2013,WeirIOP2013}.  Many of the experiments involving superconducting rings were conducted exclusively in this regime, and required a very large number of experimental repetitions to make accurate measurements of the very low rates of spontaneous current formation. The limited repetition rates and lifetime of a quantum gas experiment makes it an impractical setting to explore this limit. 

\begin{figure}[!t]\centering
	\centering
	\includegraphics[width=\columnwidth]{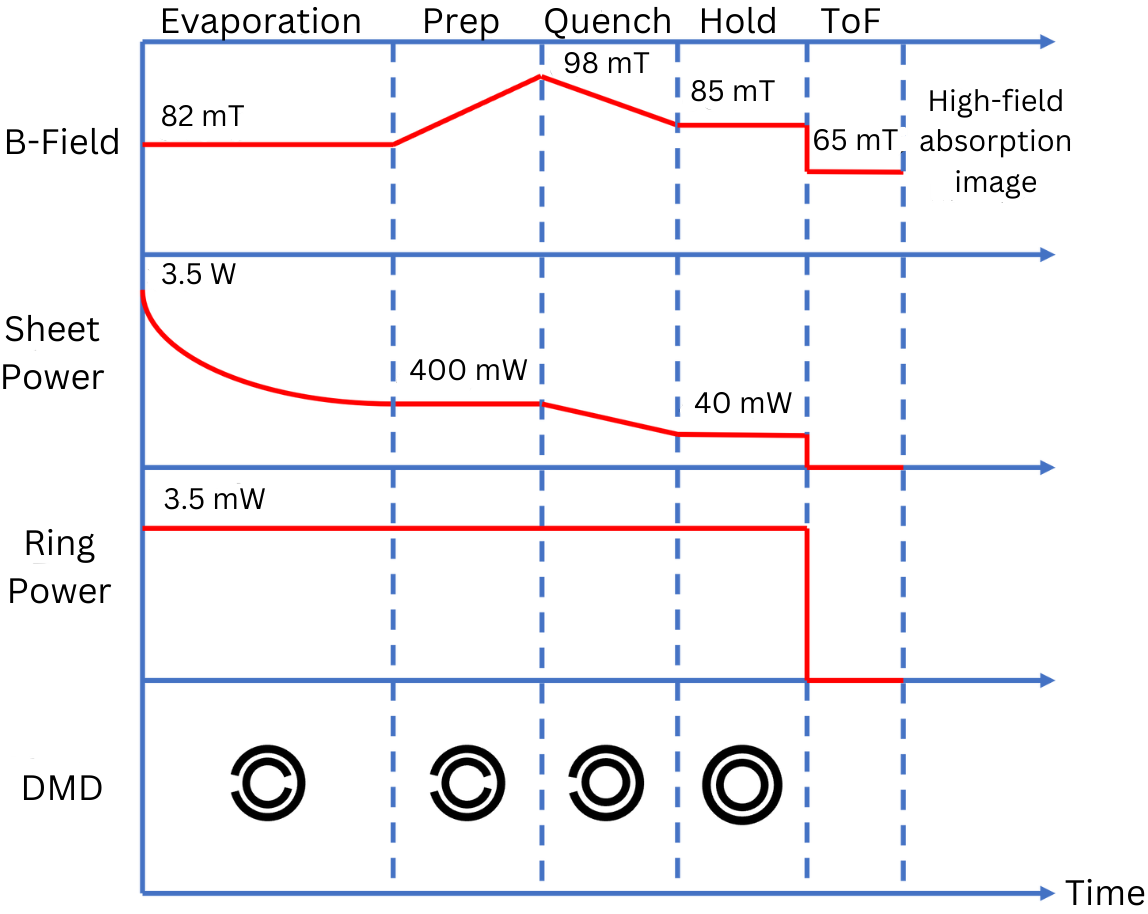}
	\caption{Schematic illustrating the sequence of stages in the preparation and hybrid quench procedure. Neither the vertical axes nor the time axis is to scale with respect to the actual experimental values. The ``Quench'' stage occurs over a variable ramp time interval $t_r$, while all of the other stages occur over fixed time intervals, and have self-explanatory labels. The state preparation stage terminates at the end of the phase labeled ``Prep", at which point the hybrid quench proceeds. The diagram in the row labeled DMD represent the time-varying pattern of the red-detuned beam used to create the concentric double-ring potential, where the black color denotes high-intensity (low potential) regions of the trap.}
	\label{fig:Schematic}
\end{figure}

\section{Experimental Methods} \label{sec:exp_methods}
The standard experimental method for quenching quantum gases through the superfluid phase transition has been to use evaporative cooling, but for atoms with a Feshbach resonance like lithium, it is also possible to drive the system through a phase transition with a ramp of the interaction strength. Our work and another experiment reported very recently~\cite{LeeArXiv2023} are the first to make use of interaction ramps in studying the KZM in an ultracold quantum gas.

\begin{figure}[!t]\centering
	\centering
	\includegraphics[width=\columnwidth]{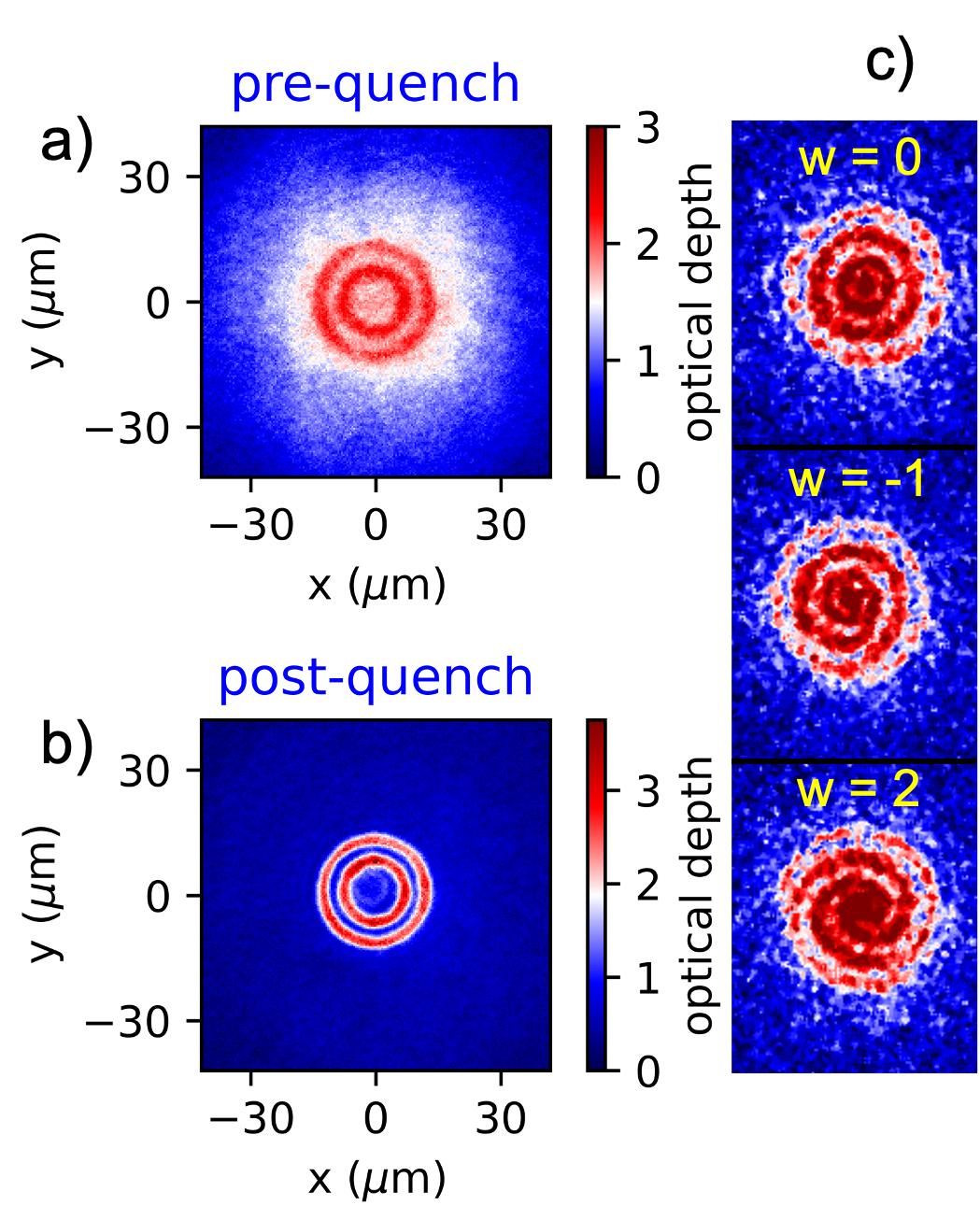}
	\caption{Interferometric detection of persistent currents. a) Average optical depth of 20 absorption images showing the pre-quench state of the system, with an equal mixture of $N=9.6\times 10^4$ total atoms in a double-ring potential, at a field of $97.6$ mT and $T/T_C\approx2.7$. b) Average optical depth of 20 images taken after a $50$ ms hybrid quench (see text), with $N=6.4\times 10^4$ atoms remaining. The broad, dilute thermal halo is present in both images, and is responsible for limiting the deleterious effects of heating and atom loss due to collisions, and for maintaining a roughly constant peak density in the ring dimple region during the quench. A small, dilute ring-shaped region of atomic density within the inner ring can be seen, and is likely due to ghosting of the ring beam generated from reflections off the many optical surfaces in the projection and imaging assembly. c) Examples of distinct matter-wave interference patterns for different winding numbers, which appear in absorption images taken $1.3$ ms after the trap potential is shut off (single realizations). The optical depth color bars are not shown for these images.}
	\label{fig:In-situ KZM}
\end{figure}

The details of the trap and the pre-quench state preparation are outlined schematically in Fig.~\ref{fig:Schematic} and Fig.~\ref{fig:In-situ KZM}(a). Additional technical information can also be found in Refs.~\cite{CaiPersistentPRL2022, AllmanPRA2023}. We begin with a roughly equal mixture of $10^5$ total $^6$Li atoms in the two lowest energy hyperfine states, which have broad Feshbach resonance at 83.2 mT. We prepare the system at a field of $98$ mT where the tunable s-wave scattering length is $a=-0.236$ $\mu$m (-4460 $a_0$, with $a_0$ the Bohr radius). The atoms are transversely confined in a double ring dimple geometry superimposed upon a broad sheet-like background, where the radii of the inner and outer rings are 7.5 $\mu$m and 12.5 $\mu$m, respectively, with Gaussian half-widths of 1.5 $\mu$m, as shown in Fig.~\ref{fig:In-situ KZM}a and b. The inner and outer rings are kept at the same depth throughout the experiments, and the oscillation frequency for radial motion about their minima also remains constant at $f_r=5.5$ kHz.

The initial Fermi energy determined from a model of the trap is $E_F=h\times33.2$ kHz ($h$ is Planck's constant), and the interaction parameter $1/k_Fa\approx -0.7$ in the regions of highest density. (See Table~\ref{table:table_of_numbers} for further details.)
We obtain the initial system temperature from measurements of the density profile of atoms which form a dilute non-degenerate ``halo" with a radius significantly larger than the outer ring and find that $(T/T_F)_{\text{initial}}\approx 0.25$ in the sheet \cite{Chevy2022,AllmanPRA2023}.  With the critical temperature given by $T_c/T_F=0.277\exp(-\pi/2k_F|a|)$ \cite{Gorkov1961}, we confirm that the system is entirely in the normal phase with $(T/T_c)_{\text{initial}}\approx 2.7$. The halo serves an additional desirable purpose of mitigating collision-induced heating and atom loss from the ring region during quenches \cite{AllmanPRA2023}. 

To initialize the system in a non-rotating state before the quench, we raise narrow barriers in both rings over $100$ ms. This is achieved by updating the pattern on the digital micromirror device (DMD, Texas Instruments DLP Lightcrafter 6500) controlling the vertically-propagating red-detuned beam used to generate the potential. The barrier in the inner ``experiment'' ring is then lowered over the same time scale, reconnecting the ring before the quench. In contrast, the barrier in the outer ``reference'' ring is maintained until after the quench, but lowered before the atoms are released from the trap for interferometric detection (see Fig.~\ref{fig:Schematic}). This procedure ensures that the reference ring is in a zero-current state when the atoms are released from the trap and allowed to interfere. We selected the inner ring as the ``experiment'' ring because optical imperfections more strongly affected the uniformity of the outer ring.

\begin{table}[t]
\renewcommand{\tabularxcolumn}[1]{m{#1}}
\begin{tabularx}{\columnwidth}{XXXX}
\hline\hline 
   Parameter                & Initial  & Critical & Final \\
   \hline
   $t/t_r$                  & 0        & 0.61     & 1    \\
   $B$ (mT)                 &  98      &  90      &  85  \\
   $P_\mathrm{s}$ (mW)      &  400     &  180     &  40  \\
   $f_z$ (kHz)              &  4.7     &  3.2     &  1.5 \\
   $a$ ($\mu$m)             & -0.236   &  -0.408  & -1.31\\
   $N$ ($10^4$)             & 10       & 7.8      & 6.4  \\
   $T_c$ ($\mu$K)           & 0.17     & 0.19     & 0.22 \\
   $T_F$ ($\mu$K)           & 1.6      & 1.3      & 1.0  \\
   $1/k_F a$                & -0.70    & -0.43    & -0.15\\
   $\lambda_F$ ($\mu$m)     & 1.0      & 1.1      & 1.3  \\
   $t_F$ ($\mu$s)           & 30       & 36       & 48   \\
   $T$   ($\mu$K)           & 0.40     & 0.19     & 0.05 \\
   $T/T_F$                  & 0.25     & 0.14     & 0.05 \\
   $T/T_c$                  & 2.7      & 1        & 0.24 \\
   $\epsilon$               & -1.7     & 0        & 0.76 \\
   $t_q/t_r$                & 0.31     & 0.43    & 0.63  \\
\hline\hline
\end{tabularx}
\caption{Controlled, measured and calculated system parameters at key points during the simultaneous ramp of magnetic field $B$ and sheet beam power $P_s$. We additionally include the total atom number $N$, Fermi wavelength $\lambda_F=2\pi/k_F$, and Fermi time $t_F=h/E_F$.}
\label{table:table_of_numbers}
\end{table}

After preparing the system with a connected inner ring and broken outer ring, we drive it into the superfluid phase with a hybrid quench, where both the temperature $\textit{and}$ the critical temperature are ramped simultaneously over a variable ramp time $t_r$. Temperature is reduced by forced evaporative cooling, achieved by lowering the sheet beam power linearly from $400$ mW to $40$ mW. The critical temperature is increased by lowering the magnetic field linearly from $98$ mT to $85$ mT, which changes the scattering length from $a$ = -0.236 $\mu$m to -1.31 $\mu$m (-4460 $a_0$ to -24800 $a_0$). We chose to use this approach because we found that we could achieve faster effective quench rates while using less extreme ranges of control parameters. 

We obtained measurements of the final current state of the inner ring for ramp times ranging from $50$ ms to $4.4$ s, with the minimum time set by limits on the slew rate of the magnetic field. While our vacuum-limited lifetime allowed for quenches longer than $4.4$ s, we did not observe any non-zero spontaneous currents in this limit. For the ramp parameters given above, the system is predicted to go through the superfluid transition 60$\%$ of the way through the ramp, for all ramp durations. Additional information about the controlled, measured, and calculated system parameters during the quench are provided in Table~\ref{table:table_of_numbers}. The freeze-out dynamics occur over a sufficiently small part of the ramp  ($\hat{\epsilon}\lesssim0.01$, See Sec.~\ref{Results and Discussion}) that it is still reasonable to treat the change in reduced temperature as linear over that range, even though the scattering length and critical temperature are nonlinear functions of the magnetic field, especially toward the end of the ramp. Furthermore, due to the tight transverse confinement of the rings, we can safely approximate the critical temperature by its peak value, and in contrast to experiments performed in broad harmonic traps disregard modifications of the quench dynamics due to the inhomogeneity in the transverse directions. For the conditions in our experiment, the ramp duration $t_r$ and the instantaneous quench time $t_q(t)\equiv 1/\dot{\epsilon}(t)$ at the critical point are related by $t_q\approx 0.43t_r$, where we emphasize that $t_q(t)$ is time-dependent due to the weak non-linearity of the ramp.

After the end of the quench we remove the barrier in the outer ring adiabatically over $100$ ms to allow matter-wave interference with the non-rotating reference ring. We then simultaneously turn off the trapping beams and the current in one pair of bias magnet coils. The rapid magnetic field jump to 65 mT efficiently converts the weakly-bound Cooper pairs to molecules~\cite{RegalPRL2004} which is essential for preserving the coherence of the pairs during ballistic expansion. After $1.3$ ms we measure the molecule density using resonant absorption imaging of atoms in the $|1\rangle$ state on a closed transition at $65$ mT. The winding number of the phase around the inner ring is clearly visible as a spiral in the pattern created by matter-wave interference between the inner and outer ring, as shown in Fig.~\ref{fig:In-situ KZM}(c).

In testing our experimental protocol we also considered the possibility of post-quench decay of the persistent currents before detection. To assess the rate of decay (via e.g. thermally activated phase slips) we repeatedly initialized the inner ring (with approximately $100\%$ fidelity) in the highest observed $|\ell|=2$ persistent current state using a blue-detuned ``stirring" beam~\cite{CaiPersistentPRL2022}. We observed no instances of current decay even for hold times exceeding $5$ s, and concluded that the decay rate was negligible.

We also considered the possibility of simultaneously quenching two unbroken concentric rings, which can boost the dynamic range of the winding number difference while preserving the essential scaling laws in the KZ regime of fast quenches. This occurs if the winding number distributions for each ring are statistically independent, each possessing a variance that scales as a power law of the quench time in the KZ regime. Interestingly, the Zurek argument suggests the winding number difference between two independent rings of circumferences $C_1$ and $C_2$ should act statistically equivalent to the winding number in a single ring of circumference $C_1+C_2$. Evidence of spurious rotational biases in exploratory work on unbroken rings, however, motivated us to develop the procedure using barriers described above, which greatly reduced the asymmetry of the current distributions.

\begin{figure}[!t]\centering
	\centering	\includegraphics[width=\columnwidth]{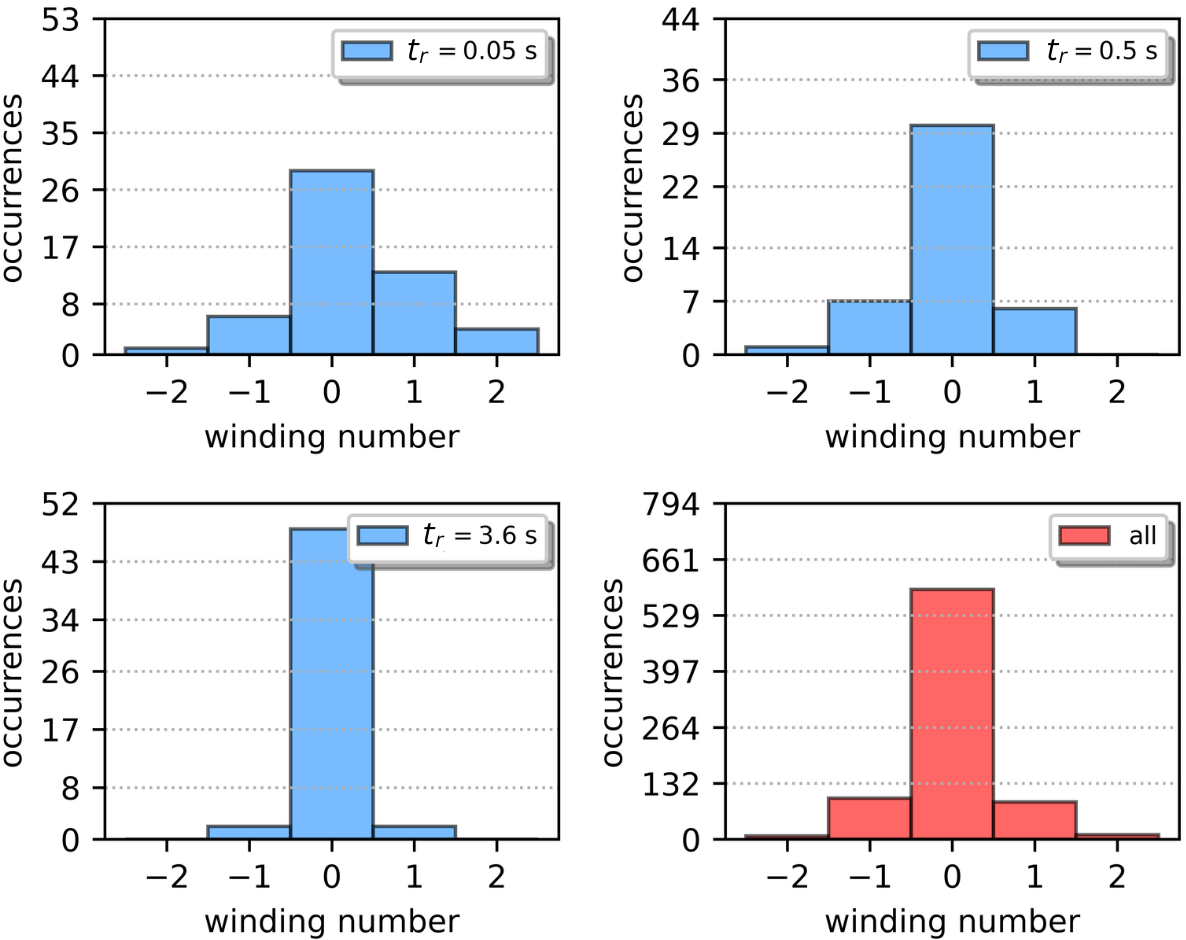}
	\caption{Observed occurrences of winding numbers for various quench times (blue histograms). Each histogram contains at least 40 samples. The number of samples for each displayed histogram is given by the largest number on the vertical axis. We also show a histogram of winding numbers for all quenches (red histogram). The average of all winding number measurements combined $\langle w\rangle_{\text{all}}= 0.0(2)$, indicating that there are minimal biases to the winding number distribution. The uncertainty is computed as the standard error of the mean of all winding numbers. The asymmetries in the histograms for all quench rates except the fastest one are well within the limits of sampling error. The p-value for a shift from a zero-mean distribution for the fastest quench is 3.2$\%$, and weakly indicative of a bias from bulk flow.}
	\label{fig:KZM histogram}
\end{figure}

\section{Results and Discussion}\label{Results and Discussion}

We repeated the experimental procedure described above a minimum of 40 times for 17 different quench rates, sampling the distribution of winding numbers over the entire experimentally accessible range.  Several example histograms of measured winding numbers are shown in Fig.~\ref{fig:KZM histogram}. We found that the distribution of measured winding numbers was reliably peaked around $w=0$ as long as the barriers were applied as described to eliminate any bias arising from currents in the normal state or the reference ring, with only the fastest quench still showing some small indication of bias due to bulk flow that was possibly induced during the quench. 

\begin{figure}[!t]\centering
	\centering
	\includegraphics[width=\columnwidth]{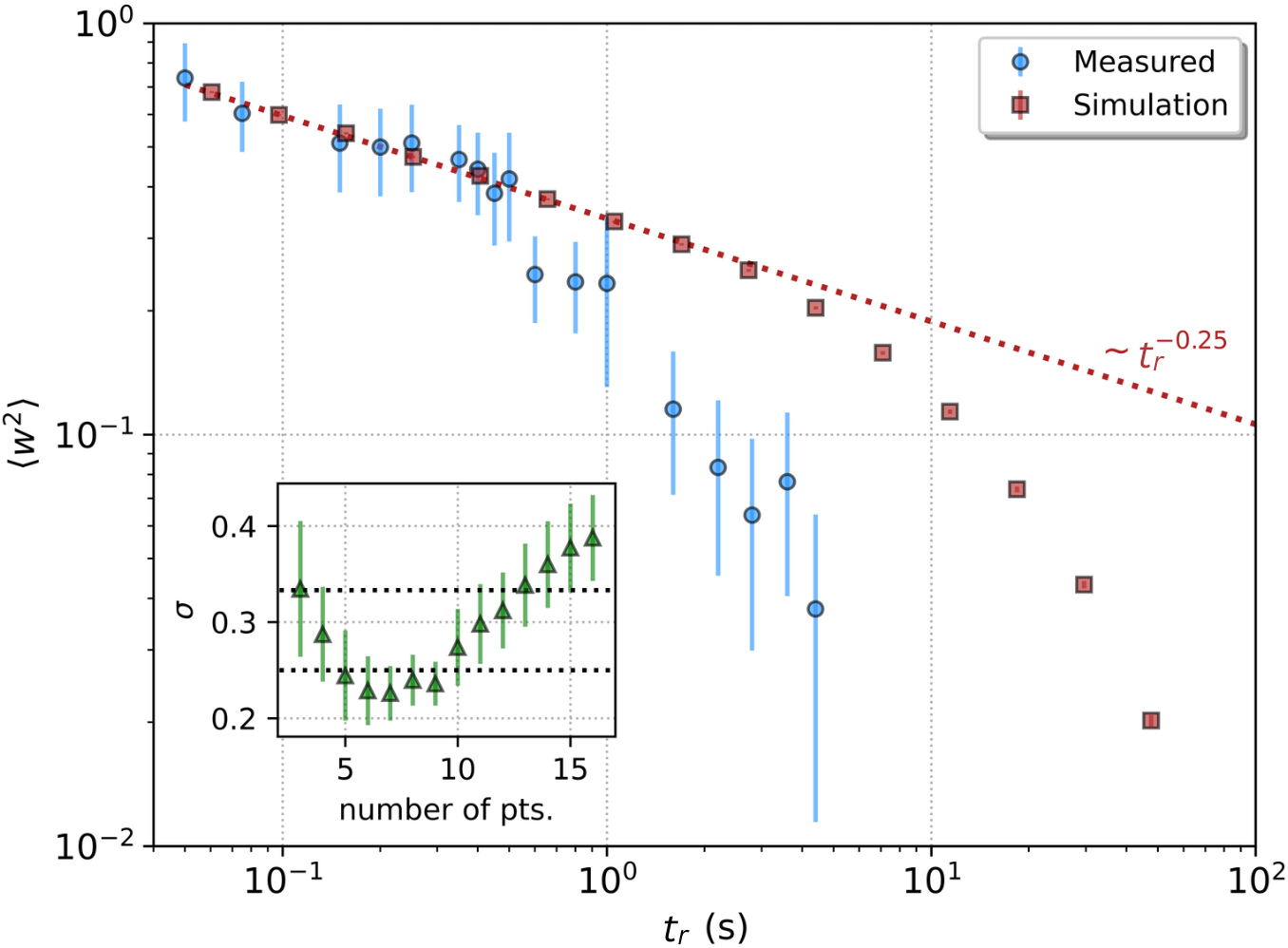}
	\caption{Plot of measured mean-square winding number versus ramp duration (blue circles). A power-law fit to the measured data for the nine fastest quenches yields a scaling exponent $\sigma=0.24(2)$. We also show the variances obtained from the simulated winding number distribution obtained using the 1D SLGE (red squares). The straight red dotted line shows a power law with exponent $0.25$. The inset shows the power law fit-extracted exponents $\sigma$ obtained from fits to various numbers of fastest-quench data points. The dotted lines show the mean-field and F-model predictions $\sigma=1/4$ and $1/3$, respectively.}
	\label{fig:KZM expt plot}
\end{figure}

We show in Fig.~\ref{fig:KZM expt plot} the mean-square winding number of spontaneous currents for all the quench rates sampled in this experiment. The vertical error bars are 1-$\sigma$ uncertainties obtained from a bootstrapping technique employed because the discreteness of the winding number distribution is important under our experimental conditions \cite{johnson2001introduction}. Values of the mean square winding number for our fastest quenches are consistent~\cite{Aidelsburger2017} with an average number $N_d\sim10$ domains formed at freeze-out. For the nine fastest quenches, we see an approximate power law scaling of the form $\langle w^2\rangle\sim t_r^{-\sigma}$ with an exponent falling in a range between about 0.2 and 0.3. However, the best-fit value shifts appreciably when the points included in the fit are varied, as shown in the inset of Fig.~\ref{fig:KZM expt plot}. (Uncertainties are a 1-$\sigma$ confidence interval obtained from the fit covariance matrix). Using the first five to nine points gives a roughly consistent exponent of $0.24(2)$, which is in line with the scaling expected from mean-field predictions ($\nu_{MF}=1/2$ and $z_{MF}=2$ give $\sigma_{KZ}=1/4$).

However, mean-field scaling is only expected to occur if the freeze-out occurs far enough from $T_c$ that the fluctuations of the order parameter are smaller than its mean~\cite{larkin2005theory, deMeloPRL1993, LobbPRB1987}, and our system is strongly interacting. The range of reduced temperatures where beyond-mean-field physics is expected to be important is given by the Ginzburg-Levanyuk number, $\mathrm{Gi}$. In a uniform three-dimensional (3D) system of strongly interacting fermions, $\mathrm{Gi}\sim$ 0.2 at $1/k_Fa=-0.4$~\cite{Taylor2009, Pascucci2021}. In the KZM framework, the reduced temperature at the freeze-out time should be $\hat{\epsilon}=(\tau_0t_q^{\nu z})^{1/(1+\nu z)}$, where $\tau_0$ is on the order of the Fermi time $t_F\approx35$ $\mu$s. From these relations we expect $\hat{\epsilon}\sim 0.001$ for our slowest quenches and $\sim 0.01$ for the fastest, placing the freeze-out in the critical region where beyond mean-field effects are significant. 

Under these conditions, the predicted scaling exponents are those of the 3D XY model (sometimes referred to as the F model \cite{HohenbergRMP1977}) for which $\nu_F=2/3$, $z_F=3/2$ and $\sigma_{KZ}=1/3$. There is some tension between our data and this prediction, and further experimental work will be required to refine the measurements and better understand the factors involved. One potential issue is that ramping the interactions causes $T_c$ and $\mathrm{Gi}$ to increase during the quench, and it is possible that this is leading to unexpected effects that could be identified in future experiments comparing results from pure evaporation quenches, pure interaction quenches, and hybrid quenches. Another potential issue is that it may not be reasonable to expect our system to conform to the predictions for a uniform 3D system, given that its radial width is only a few times the Fermi length. Varying the ring width may help identify whether this is the case. 

For slower quenches, we observe a clear departure from KZ scaling into a regime where the mean-square winding number falls more rapidly with quench time. We had expected this to occur as the typical domain size becomes comparable to the circumference of the ring~\cite{WeirIOP2013} but found that the probability for spontaneous currents to appear falls off somewhat faster than predicted by a linearized stochastic Ginzburg-Landau model we used for quantitative comparison (see Sec.~\ref{PKZM}). This unexpectedly  sharp decrease in current formation in the slow quench limit could instead be the result of weak coupling between the transverse acoustic modes of the two rings on longer timescales. This is possible because the local chemical potential in the region between the rings is small but nonzero at the time the high-density regions pass through the phase transition. When the time scale associated with this coupling becomes comparable to the condensate growth time scale $\hat{t}_{KZ}\sim(\tau_0t_q)^{1/2}$ (which holds for both mean-field and F-model universality classes since $\nu z=1$)  adjacent regions of the two rings may phase-lock before independent and topologically protected persistent currents can form within each ring. In this case non-circulating states in the experiment ring become more probable, because the outer phase reference ring is by design encouraged to be non-circulating.

We estimate that the timescale for coupling between the rings is on the order of 1 ms, using simple circuit-hydrodynamic model where the rings are treated as a capacitance connected by the linear kinetic inductance associated with normal-state flow through the region of lower density that separates them~\cite{GauthierPRL2019}. For the ramp duration $t_r\approx 0.5$ s ($t_q\approx 0.2$ s) at which the winding number variance falls abruptly, the condensate growth time scale is also estimated to be of order 1 ms, taking $\tau_0$ to be the Fermi time at the critical point (see Table~\ref{table:table_of_numbers}). We also note that theoretical investigations of Josephson vortex formation after a quench of tunnel-coupled rings predicted qualitatively similar effects when coherent coupling between rings becomes large enough~\cite{BrandPRL2013}.  In future experiments employing a blue-detuned double ring, this effect can be studied in greater detail by tuning and even eliminating the inter-ring coupling via the local chemical potential between the rings.

We also note an important distinction between the dimensionality of thermal excitations near the ring-shaped minimum and the dimensionality of fluctuations in the order parameter as it is driven through the transition. The fermionic thermal excitations have predominantly three-dimensional character as determined by the ratios $k_BT_c/\hbar\omega_\perp$ and $E_F/\hbar\omega_\perp$, where $\omega_\perp$ is the largest transverse trap angular frequency, in this case the angular frequency ($\omega_r=2\pi\times5.5$ kHz) associated to radial motion about the ring-shaped potential minima. While the former ratio is of $\mathcal{O}(1)$ near the freeze-out point, the latter is typically of $\mathcal{O}(5)$. Transverse excitations of the order parameter, on the other hand, are strongly suppressed since the correlation length of phase fluctuations at the freeze-out time is always larger than the ring width in our quenches. Thus, although the phase transition is probably best described as three-dimensional in character, the dynamics of the order parameter near the transition should have effectively one-dimensional character. 

Finally, we briefly discuss possible extensions to this work that are within experimental reach. First, as was pointed out in~\cite{McDonaldPRA2015}, Kibble-Zurek scaling laws may be modified by reservoir interactions. In particular, energy-exchanging collisions between the coherent population comprising the order parameter and the surrounding incoherent thermal bath were shown to significantly increase the dynamical scaling exponent $z$. This in turn leads to a decrease in the KZ exponent of the winding number variance. A natural extension of this work would be to study scaling laws in a trap where any non-degenerate fermionic ``halo," if present, can be selectively isolated from the double ring. Experimentally this would be most easily achieved by employing a blue-detuned ring trap, and in this way, reservoir interactions and their effects on KZ scaling laws may be studied. As mentioned earlier, employing a blue-detuned ring trap would have the additional advantage of decoupling possible inter-ring interactions during slower quenches. 

A second extension of this work regards the potentially interesting role that fermionic pairing may play in the generation of spontaneous currents following quenches across a pair-superfluid phase transition. It is known that there is a temperature gap separating superfluid transition temperature $T_c$ and the temperature at which bound fermion pairs form $T^*$ (see, e.g.,~\cite{BOETTCHER201263}). For transitions closer to unitarity, it is a natural question to ask what role so called pre-formed pairs --non-condensed yet bound fermionic pairs that exist in a substantial temperature range $T_c<T<T^*$-- play on the formation of spontaneous currents. Deviations from the usual KZ scaling laws, however, have not yet been seen for quenches covering a modest range of the BEC-BCS crossover~\cite{Ko2019}. However, for transitions taking place in the deep BCS limit, the temperature band in which these pre-formed pairs can exist approaches zero. Thus, one would expect that the statistics of spontaneous current formation would in this limit be affected by the rate at which unbound fermions can bind into Cooper pairs and subsequently attempt to establish macroscopic order as the transition is crossed, as opposed to the usual KZ scenario which presupposes the existence of some fluctuating yet disordered coherent field above the transition. With modifications to our magnet coil circuitry that enable controlled field jumps from the BCS limit to the near-unitary limit on $10-100$ $\mu$s time scales, the role of this fermionic pairing mechanism on the KZM can be studied. This would be a complimentary experiment to those performed in~\cite{DykePRL2021, Liu2021}, which directly investigated the rate of pair formation and condensate growth after rapidly crossing the transition on the BCS side of the Feshbach resonance.

\section{Phenomenological KZM}\label{PKZM}

We now provide an analytical framework for reproducing and interpolating between the fast and slow quench limits in the absence of inter-ring coupling. To do this, we treat the Fourier components of the fluctuating order parameter as Gaussian variables evolving according to an overdamped, stochastically-driven Landau-Ginzburg model. In a single smooth quasi one-dimensional ring of radius $R$, the time-dependent stochastic Landau-Ginzburg equation (SLGE) describing the evolution of the order parameter can be written 
\begin{equation}\label{SLGE}
    \frac{\partial\psi}{\partial t}=\left(\alpha(t)+\frac{\partial^2}{\partial\theta^2}+\beta|\psi|^2\right)\psi+\zeta(
    \theta,t)
\end{equation}
which is expected to approximate the dynamics at the mean-field level~\cite{DasWindingSR12}. Here, we measure time in units of $\gamma/\Omega_0$, where $\gamma$ is a phenomenological dimensionless relaxation rate and $\Omega_0=\hbar/m_pR^2$ is the frequency associated with the quantized circulation of pairs of mass $m_p$ around the ring. In addition, $\alpha(t)=(R/\xi_{\text{BCS}})^2\epsilon(t)$ is the dimensionless Landau-Ginzburg chemical potential, written in terms of the BCS coherence length $\xi_{\text{BCS}}$, and $\beta$ is the non-linear interaction strength. Further, $\zeta(\theta,t)$ is a zero-mean complex Gaussian white noise field, satisfying $\langle\zeta(\theta,t)\rangle=0$ and $\langle \zeta^*(\theta,t)\zeta(\theta',t')\rangle=D\delta(\theta-\theta')\delta(t-t')$, with $D\sim k_BT$ a phenomenological ``diffusion" constant. Finally, angular brackets denote averaging over the independent and identically distributed normal distributions from which the $\zeta$ are pulled.

To study quench dynamics, we vary the reduced temperature linearly as $\epsilon(t)=t/t_{q,0}$ (distinguishing the dimensionfull quench time $t_q$ from the dimensionless quench time $t_{q,0}=\Omega_0t_q/\gamma$), and we can write $\alpha(t)\equiv t/t_\alpha$ with $t_\alpha\equiv(\xi_{\text{BCS}}/R)^2t_{q,0}$. Close to the transition, where the length scale of fluctuations in the order parameter is expected to become frozen in, we may neglect the non-linear term and write a linearized Fourier space representation of \eqref{SLGE}:
\begin{equation}\label{LSLGE}
    dc_{\ell}/dt=(t/t_\alpha-\ell^2)c_{\ell}+\zeta_{\ell}(t)
\end{equation}
where $\psi(\theta,t)=\sum_\ell c_\ell(t)\exp(i\ell\theta)$ and $\zeta(\theta,t)=\sum_\ell \zeta_\ell(t)\exp(i\ell\theta)$. We take the quench to begin at $t=-\infty$ where $\psi(\theta,t=-\infty)=0$. The central objects essential to describing the statistics of spontaneous current formation are the mean-square fluctuations of the discrete Fourier components, obtained from the formal solution to \eqref{LSLGE}:
\begin{align}\label{MS_fluctations} 
\langle|c_{\ell}(t;\hat{t})|^2\rangle&\equiv\sigma_{\ell}^2(t;\hat{t})=\sqrt{\pi}D\hat{t}\ F(t/\hat{t}-\hat{t}\ell^2)
\end{align}
where $\hat{t}\equiv\sqrt{t_\alpha}$. The dimensionless function $F(x)\equiv \exp(x^2)[1+\text{erf}(x)]/2=\text{erfcx}(-x)/2$ where $\text{erfcx}$ is the complimentary scaled error function. At any time $t$, $\sigma_{\ell}^2(t;\hat{t})$ is symmetrically peaked around $\ell=0$. Additionally, the growth dynamics of the $\sigma_{\ell}^2$ depend only on the variable $x_{\ell}(t)\equiv t/\hat{t}-\hat{t}\ell^2$. Since $F(x)\sim\exp(x^2)$ for $x\gtrsim 1$, the fluctuations in mode $\ell=0$ experience a brief period of rapid growth, before any other mode, following the transition at times $t\approx\hat{t}$ when $x_{\ell=0}(t)\approx 1$. For $t\gtrsim\hat{t}$, non-linear effects kick in and the condensate begins to relax toward its instantaneous, non-zero equilibrium value~\cite{Liu2020}. Thus, in some short interval of time following this ``blow-up" time $\hat{t}$, the condensate becomes robust with respect to fluctuations large enough to cause any persistent current to decay; the winding number becomes a topologically protected quantity at times $t\geq t_{\text{eval}}\equiv f\hat{t}$. Here $f>1$ is a roughly constant ``fudge" factor that scales the blow-up time to the so-called evaluation time where the winding number is stabilized \cite{CheslerPRX2015}. 

The goal now is to calculate the probability of observing a given winding number given the set of time-dependent, Gaussian random Fourier coefficients $\boldsymbol{c}(t)$ obtained from \eqref{LSLGE}. At any given time the winding number $w$ can be  obtained from the density-phase representation of the order parameter $\psi(\theta)=\sqrt{n(\theta)}\exp[i\phi(\theta)]$. To compute $w$ directly, we may logarithmically differentiate this expression [assuming $n(\theta)>0$, which is almost certainly true at any given time] and then integrate around the ring, using the definition $\int_0^{2\pi}d\theta d\phi(\theta)/d\theta=2\pi w$ and then the substitution $z=\exp(i\theta)$:
\begin{equation}\label{winding number}
\begin{split}
    w&=\frac{1}{2\pi i}\int_0^{2\pi}d\theta\frac{d}{d\theta}\ln\psi(\theta)\\
    &=\frac{1}{2\pi i}\oint_{|z|=1}dz\frac{d}{dz}\ln{\Psi}(z;\boldsymbol{c})
\end{split}
\end{equation}
where we have defined the Fourier-like expansion of the order parameter, truncated at modes $\ell=\pm\ell_c$, as
\begin{equation}\label{psi}
    \Psi(z;\boldsymbol{c})=\sum_{\ell=-\ell_c}^{\ell_c}c_\ell z^{\ell}.
\end{equation} 
Using Cauchy's argument principle, the winding number is then given by
\begin{equation}\label{winding}
    w=m-\ell_c
\end{equation}
with $0\leq m\leq 2\ell_c$ the number of roots of $\Psi(z;\boldsymbol{c})$ lying within the complex unit disk $|z|<1$. This Fourier-space method of computing the winding number circumvents issues with phase ambiguities associated with the real-space computation of $w$.

We numerically simulate the winding number distribution by sampling the $c_\ell(t;\hat{t})$ from the complex Gaussian distribution $\mathcal{C}\mathcal{N}(0,\sigma_\ell(t;\hat{t}))$. The variances $\sigma_\ell^2(t;\hat{t})$ are given by \eqref{MS_fluctations}. For a single randomly-chosen set $\boldsymbol{c}(t;\hat{t})$, the roots of $\Psi(z;\boldsymbol{c}(t;\hat{t}))$ are found numerically and $w(t)$ is then computed via \eqref{winding}. When evaluated at $t_{\text{eval}}$, the probability distribution depends only on the variable $t_\alpha\sim t_q/R^4$ where the temporal dimension was restored. Notably, the phenomenological diffusion constant $D$ drops out of the winding number distribution as long as the winding numbers assume their final values at $t_{\text{eval}}$ and non-linear effects can be neglected. We note that the algebraic decay of the $\sigma_{\ell}^2\sim 1/\ell^2$ at large $|\ell|$ means that a large $\ell_c$ is required to faithfully simulate the winding number distribution, especially for fast quenches. We chose a generous $\ell_c=1000$ for simulations and confirmed that our results across all quench times did not change significantly after increasing this value. 

We show in Fig.~\ref{fig:KZM expt plot} the numerically simulated mean-square winding number evaluated at dimensionless time $t_{\text{eval}}=3\hat{t}$. To compare this theoretical prediction with measurements without knowledge of phenomenological model parameters, we rescaled the model quench time $t_\alpha$ to best fit the data in the fast quench regime. We see good agreement in the measured and simulated data for the fastest quenches, bolstering the Zurek argument for spontaneous current formation in 1D rings. Additionally, we see a rapid fall-off in the mean-square winding number for slower quenches in both the measured and simulated data, although the simulated data do not match the measured values in this regime. In the model, this fall-off is attributed to finite-size effects, although the rate of fall-off depends on microscopic parameters and ``fudge factors" that are difficult to estimate experimentally. As discussed previously, it may be necessary to eliminate the effects of ``phase-locking" between the rings with an impenetrable barrier in order to explore this regime.

\section{Conclusion}\label{conclusion}

We have studied the statistics of spontaneous current formation in a thermally-quenched ring of ultracold fermionic atoms with strong attractive interactions, where the currents are long-lived and topologically protected. We observe a fast-quench regime where the mean-square of the winding number follows an algebraic scaling as expected from KZ theory, and a slow-quench regime governed by a more rapid suppression of spontaneous current formation. A stochastic Landau-Ginzburg model of spontaneous current formation showed quantitative agreement with data for fast quenches but only qualitative agreement for slow quenches that we attributed to finite size and possible inter-ring coupling effects. It is somewhat unexpected that the observed scaling exponent of 0.24(2) for fast quenches is better in line with the value of 1/4 obtained from mean-field theory than the value of 1/3 predicted for the 3D XY model, given our estimates of how strong fluctuations should be at the freeze-out time under the conditions of our experiment. There are many complications involved in real experiments that can impact the observed scaling, and it appears that further work will be required to determine how the scaling might be affected by factors like using a fast interaction ramp, being close to a dimensional crossover, and having two rings coupled by a dilute normal-state background.

\begin{acknowledgments}
We thank Roberto Onofrio and Rufus Boyack for numerous helpful discussions. This work was supported by the National Science Foundation (Grant No. 2046097).
\end{acknowledgments}

\bibliography{KZM}

\end{document}